# The critical role of hot carrier cooling in optically excited structural transitions

Wen-Hao Liu[1,2], Jun-Wei Luo[1,2,3]*, Shu-Shen Li[1,2,3], and Lin-Wang Wang[4]*

[1]*State Key Laboratory of Superlattices and Microstructures, Institute of Semiconductors, Chinese Academy of Sciences, Beijing 100083, China*

[2]*Center of Materials Science and Optoelectronics Engineering, University of Chinese Academy of Sciences, Beijing 100049, China*

[3]*Beijing Academy of Quantum Information Sciences, Beijing 100193, China*

[4]*Materials Science Division, Lawrence Berkeley National Laboratory, Berkeley, California 94720, United States*

*Email: jwluo@semi.ac.cn; lwwang@lbl.gov

**Abstract:**

The hot carrier cooling occurs in most photoexcitation-induced phase transitions (PIPTs), but its role has often been neglected in many theoretical simulations as well as in proposed mechanisms. Here, by including the previously ignored hot carrier cooling in real-time time-dependent density functional theory (rt-TDDFT) simulations, we investigated the role of hot carrier cooling in PIPTs. Taking IrTe$_2$ as an example, we reveal that the cooling of hot electrons from the higher energy levels of spatially extended states to the lower energy levels of the localized Ir-Ir dimer antibonding states strengthens remarkably the atomic driving forces and enhances atomic kinetic energy. These two factors combine to dissolute the Ir-Ir dimers on a timescale near the limit of atomic motions, thus initiating a deterministic kinetic phase transition. We further demonstrate that the subsequent cooling induces nonradiative recombination of photoexcited electrons and holes, leading to the ultrafast recovery of the Ir-Ir dimers observed experimentally. These findings provide a complete picture of the atomic dynamics in optically excited structural phase transitions.

The photoinduced phase transitions (PIPTs) have become an appealing approach exploring ultrafast change and manipulation of material properties owing to the recent advances in ultrafast time-resolved diffraction techniques, combining ultrafast temporal manipulation with atomic-scale spatial resolution. [1-22] The optical excitation induces a nonequilibrium occupation of excited electronic states, which could lead to periodic lattice distortions (PLDs), expose the transient metastable states [23,24], or yield a controllable phase transition to the desired phase for practical applications [5,9,10,21]. By means of photoexcitation, ultrafast phase transitions have been realized in quasi-one-dimensional (1D) [5,6,10,16,17], two-dimensional (2D)[3,7,15,20,22] and three-dimensional (3D) [9,12,18,19,21] systems. One commonly used picture to explain the phase transition is the following: the photoexcited occupation of higher electronic states modifies the energy landscape substantially so that the original metastable phase becomes now a lower energy stable phase than the original ground state phase, causing the phase transition dynamically along the potential energy surfaces (PESs). [5,7,9,25] An alternative, yet related, the picture is that the transient change in the PESs results in a non-thermal excitation of soft phonon modes, which leads to a critically damped nuclear motion following these soft phonon modes to the end phase of the PIPT. [5,25] Our previous work[26] has also pointed out that the atomic forces for driving the PIPT in $IrTe_2$ arising from occupation of the Ir-Ir dimer antibonding (bonding) states by optically excited electrons (holes). [26] All these proposed explanations consider the photoexcited carriers as the cause for the change of PESs or the generation of additional atomic driving forces but disregard completely the phenomena of their relaxation to lower energy levels that occurred within the first hundred femtoseconds after photoexcitation.

Evidence has accumulated that hot carrier cooling may play an essential role in photoinduced ultrafast structural phase transitions [7,8]. For instance, Ideta *et al.* attributed the observed ultrafast recovery of Ir-Ir dimers following their dissolution to carrier recombination.[7] Monney *et al.* postulated that partial PIPT is driven by a transient increase of the lattice temperature following the hot carrier cooling in picoseconds.[8] We have also demonstrated that one can control the structural phase transitions by selectively exciting photocarriers into designated excited electronic states. [26] The hot carrier cooling will undoubtedly alter the carrier's occupation to the excited states, thus changing the atomic driving forces. Besides, there are many possible effects in a PIPT, including electron-electron and electron-phonon interactions and thermal fluctuations. The lack of real-space atomic snapshots and the inability to turn on and off physical effects (e.g., electron-electron and electron-phonon interactions) experimentally renders it challenging to disentangle these interweaving physical effects relying on experimental measurements alone. Here, by including the previously ignored hot carrier cooling effect, we have advanced the real-time time-dependent density functional theory (rt-TDDFT) simulations to study the dynamical processes of PIPT.

In contrast, most current rt-TDDFT simulations can only describe the excited systems' immediate dynamics following the photoexcitation since they cannot describe the hot carrier cooling effect. Furthermore, in many simulations, optically excited electrons are mimicked by manually taking electrons from the top of the valence band and placed at the bottom of the conduction band, or thermally distributed using a very high temperature. [5,9,10,25,27,28] Such simulations may provide a qualitative picture for PIPT, but it is unfeasible to reproduce the transition times behavior observed in the experiments accurately. An improvement has been made in some recent rt-TDDFT simulations using a Gaussian-envelope laser pulse to represent the actual laser light. [24,26,29] But the use of Ehrenfest dynamics in the rt-TDDFT is unlikely to describe the carrier cooling correctly since it lacks the detailed balance. As a matter of fact, the Ehrenfest dynamics tend to overheat the electronic subsystem. In this work, to include the hot carrier cooling effect, we have improved our rt-TDDFT algorithm by including a Boltzmann factor, [30] which can restore the detailed balance between various electronic state transitions and hence treat the carrier cooling properly.

Figure 1 shows that after including the hot carrier cooling effect, the TDDFT simulation can precisely reproduce the experimentally measured femtosecond electron diffraction (FED) curve responsible for the complex phase transitions of $IrTe_2$ (including dissolution and recovery of Ir-Ir dimers) following photoexcitation. More specifically, the inclusion of the hot carrier cooling process in the state-of-the-art rt-TDDFT simulation significantly accelerates the phase transition within 300 fs compared to the usual rt-TDDFT simulations, which display no transition throughout the 1.2 ps simulation time. We will also explicitly illustrate that Ir-Ir dimers' ultrafast recovery following their dissolutions is caused by electrons' continuous cooling passing through the Fermi level towards nonradiative recombination with photoexcited holes. These results demonstrate that hot carrier cooling plays a vital role in the photoexcitation-induced structural phase transition.

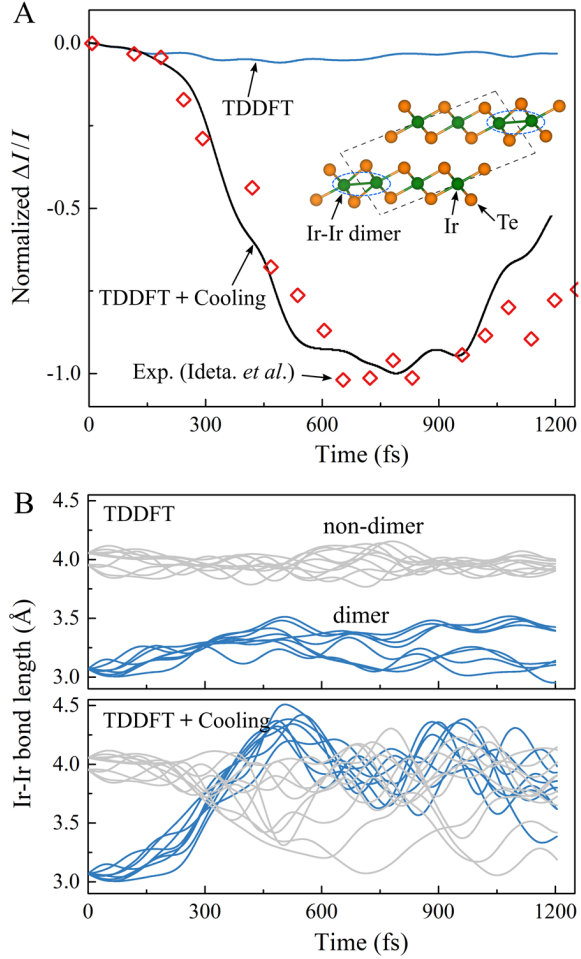

**Fig. 1. Evolution of diffraction intensity and bond length.** (A) Simulated PLD diffraction intensity $I(t)$ is obtained through the Debye-Waller formula [31-33], $I(t) = \exp[-Q^2 \langle u^2(t)\rangle/3]$, where $Q$ is the magnitude of the reciprocal lattice vector for the reflection probed, and $u^2(t)$ is average mean-square atomic displacements, as shown in Fig. S1. Red diamonds show relative diffracted intensity changes ($\Delta I/I$) measured by FED experiments. [7] Note that the zero-point time in the experiment is labeled after photoexcitation so that the diffracted intensity is far less than 0 at $t = 0$ fs. We reinstall the experimental data at $t = -380$ fs in ref. 7 as zero-point time in our work, giving a maximum value of the diffracted intensity at $t = 0$ fs. Blue and black lines represent the photoinduced lattice dynamics without and with the carrier cooling effect, respectively. The inset shows the LT phase with Ir-Ir dimerized $q = (1/5, 0, 1/5)$ lattice modulation. [34-36] The black dashed box represents the unit cell. (B) The top (bottom) panel shows the evolution of eight-pair Ir-Ir dimers (blue lines) and ten non-dimerized Ir-Ir pairs (gray lines) in the TDDFT simulation without (with) the carrier cooling.

To reveal the mechanism underlying the hot carrier cooling, we first examine the situation neglecting the hot carrier cooling process. More specifically, a straight forward rt-TDDFT simulation is carried out for the IrTe$_2$ system irradiated by a femtosecond laser pulse (central wavelength 400 nm, pulse duration 120 fs) with its amplitude tuned so that 3% of valence electrons is optically excited from the valence band to the conduction band (same as in experiments [7]). Figure. 2A shows that immediately following the photoexcitation, only 40% of the empty Ir-Ir dimers' antibonding states are filled, and the rest of the photoexcited electrons occupy the higher energy states. In our previous work, we have demonstrated that the excited electrons occupying such higher energy states tend to suppress the LT-to-HT phase transition.[26] We then proceed with the atomic dynamics by performing the rt-TDDFT simulation without including the Boltzmann factor as usual. As expected, no significant energy transfer occurs from photoexcited electrons to the lattice, and the lattice temperature remains around 200 K throughout the simulation. The top panel in Fig. 1B shows that, within the 1.2 ps simulation time, it is unlikely that the Ir-Ir dimers will undergo dissolution to achieve the LT-to-HT phase transition, which is quantified by the increase of Ir-Ir dimer bond length from 3.1 to 3.9 Å. One reason is that the photoelectrons fill only 40% of the Ir-Ir dimer's antibonding states, which is impossible to produce a strong enough atomic force to drive the phase transition.

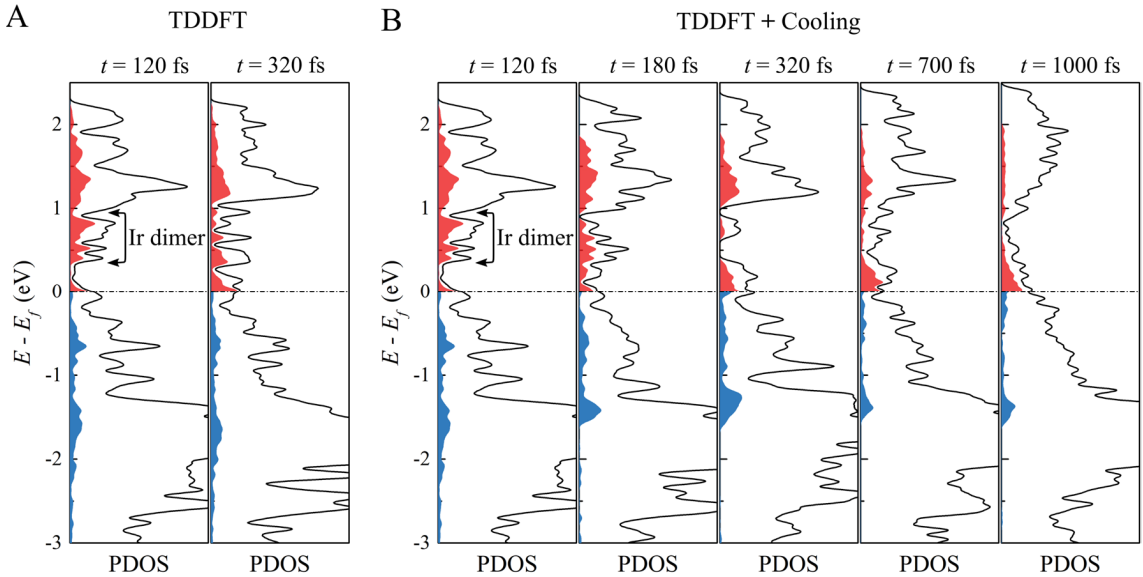

**Fig. 2. Evolution of excited carriers.** Partial density of states (PDOS) of eight-pair Ir-Ir dimers from TDDFT simulation without carrier cooling (A) and TDDFT simulation with carrier cooling (B). Red (blue) shaded areas represent 3% electronic occupations (hole occupations) where 40% Ir-Ir dimerized antibonding states are occupied by photoexcited electrons at $t = 120$ fs.

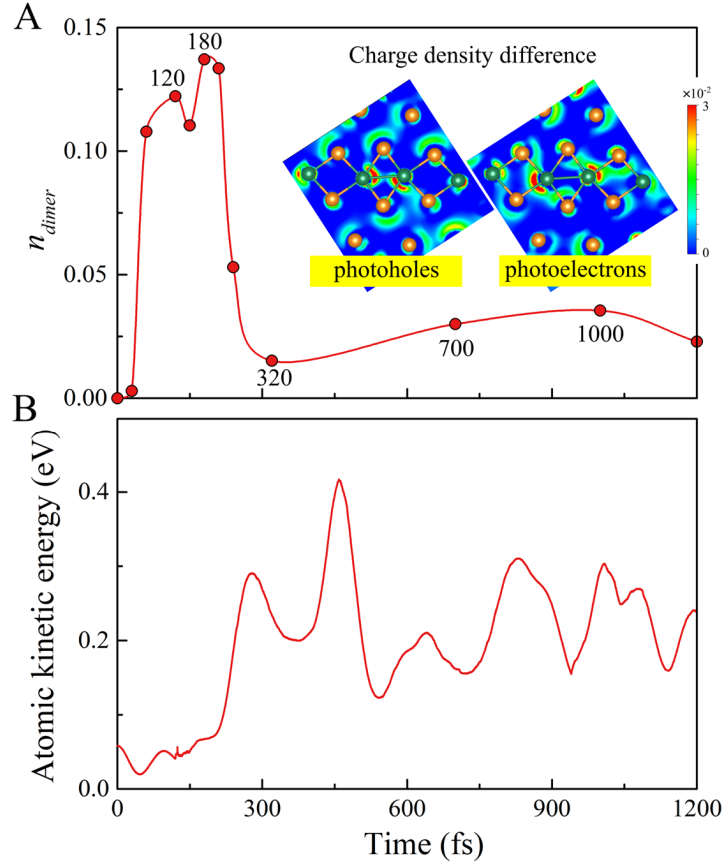

**Fig. 3. Ir-Ir dimer antibonding occupations and kinetic energy.** (A) Evolution of the electronic occupations on Ir-Ir dimerized antibonding states ($d_{xz} + d_{yz}$) within 0.3-1.0 eV above the Femi level (see Fig. S3). The inset shows the real-space charge density difference in the ($11\bar{1}$) plane between the snapshots at 200 fs and 120 fs. (B) The averaged atomic kinetic energy of Ir-Ir dimers as a function of time.

In reality, hot carriers tend to relax to lower energy electronic states and give the released energy to the lattice through electron-phonon interaction. [1-3,11-15] This transfer of the energy also heats the lattice subsystem. To explore the hot carrier cooling effect, we carry out the rt-TDDFT simulation again but by adding a particular Boltzmann factor in the algorithm [30]. Figure 2B shows the dynamic evolution of excited electrons and holes following photoexcitation. It is necessary to evaluate the dynamic filling of the Ir-Ir dimers' antibonding states by electrons considering their occupation responsible for the atomic force driving the structural phase transition [26]. To quantify it, here, we define $n_{dimer}$ as the integration of the time-dependent (due to hot carrier cooling) electronic occupation of these states (located about 0.3-1.0 eV above the Fermi level as indicated in Fig. 2):

$n_{dimer} = \int_{0.3}^{1.0} \rho_{dimer}(E,t) \, dE$. Figure 3A shows the development of $n_{dimer}$: it grows at first 120 fs from zero to a saturation value (120 fs) as a result of pulse laser photoexcitation, and then declines slightly from 120-150 fs, which indicates the emerging of hot carrier cooling process. Such decline is due to a net loss of excited electrons in the Ir-Ir dimers. The excited electrons belonging to them relax to lower energy states of other atoms but get fewer electrons from higher energy states. Figure 3B shows that, during the period, the atomic kinetic energy of the Ir-Ir dimers stays at a low level, indicating the electron-electron interaction rather than phonon-assisted process dominates the carrier cooling. During this period, the Ir-Ir dimers' bond length exhibits an invisible change (Fig. 1B).

The situation changes dramatically after 150 fs. Figure 2A shows that the photoexcited hot electrons begin to relax from higher energy levels to lower energy levels. The relaxation of hot carriers is also evidenced by the increase (a bump) in the electronic occupation of the Ir-Ir dimers' antibonding states $n_{dimer}$ from 150 fs to 210 fs, as shown in Fig. 3A. It is accompanied by an increase in the Ir-Ir dimer kinetic energy (Fig. 3B), manifesting the energy transfer of hot carrier cooling to the lattice subsystem. Thus, the phonon-assisted process is predominated in hot carrier cooling. The charge density difference between the snapshots at 200 fs and 120 fs (inset in Fig. 3A) displays that the photoexcited holes tend to localize at the Ir-Ir dimers' bond center, and electrons prefer to stay at the two ends of the Ir-Ir dimers. Such localization due to hot carrier cooling manifests the characters of Ir-Ir dimers' bonding and antibonding states, respectively. It strengthens remarkably the atomic forces for dissolution of the Ir-Ir dimers. The rapid increase of the atomic kinetic energy in the Ir-Ir dimers (Fig. 3B) should also speed up the Ir-Ir dimers' dissolution by supplying enough energy to overcome any barriers. These two factors combined together render the Ir-Ir dimers undergoing an ultrafast dissociation at about 300 fs, as indicated in the bottom panel in Fig. 1B. Figure 2B and S3 show that the Ir-Ir dimers' dissolution is along with vanishing the peaks of the Ir-Ir dimers' antibonding *d*-orbital states in DOS at 300 fs. From the time at which the lattice starts to be heating up by hot carrier cooling (at 150 fs) to the time at which the phase transition is completed (at 300 fs), the time for Ir motion is less than half of the vibrational period of the LO phonon (~ 1.2 THz) of IrTe$_2$ [26]. Such ultrafast phase transition induced by hot carrier cooling rules out the mechanisms based on unperturbed lattice mode [37]. In the unperturbed coherent lattice mode, the timescale of phase transition is the half-cycle oscillation of the coherent amplitude mode. [37,38]

To verify the effect of the increase in the atomic kinetic energy of Ir dimers unambiguously, we re-do our simulation in the NVT ensemble during carrier cooling. In contrast to the above-adopted NVE ensemble, which considers the energy transfer from hot carrier cooling to the kinetic energy in the transition degree of freedom, [30] in the NVT the lattice kinetic energy is kept constant for a given initial temperature to mimic a situation for heat dissipates quickly. Figure S2 shows that in this case, the increase in the occupation of the Ir-Ir dimers' antibonding states strengthens the atomic driving forces following hot carrier cooling. However, the Ir-Ir dimers are unable to be dissociated due exclusively to the absence in the enhancement of the Ir-Ir dimer kinetic energy. Considering the lattice can only oscillate half of its vibrational period during hot carrier cooling, as discussed above, it is difficult to claim that its kinetic energy will be dissipated out as heat during that time. Thus, we believe the NVE simulation is more closed to reality. We subsequently conclude that the hot carrier cooling inducing the increase of Ir dimer kinetic energy plays a vital role in the structural phase transition.

Furthermore, the enhanced atomic forces drive the Ir-Ir dimers' dissolution to have a deterministic and coherent manner without exhibiting a significant random fluctuation. Figure 1 shows that the system completes the LT-to-HT phase transition at 300 fs; however, the maximum change in the diffraction intensities is delayed and achieved later at 750 fs as observed in both experimental measurements [7] and our theoretical simulation. Such delay has also been detected in PIPT of the CDW material 1T-$TaS_2$, [1] and is considered as the combination of the displacive excitation of highly correlated atomic motions and phonon-induced disorder contribution to the suppression of diffraction intensity.

After reaching the minimum at around 750 fs, Fig. 1a shows that the diffraction intensity change begins to rise, which was postulated as the recovery of the Ir-Ir dimerization. [7] Here, we confirm this postulation theoretically and reveal the underlying mechanism. The difference from semiconductors, $IrTe_2$ lacks a bandgap. Thus, the cooling of the optically excited electrons is unlikely to end at the bottom of the conduction band due to the phonon bottleneck as in semiconductors. Instead, excited electrons cool down continuously, passing the Fermi energy to finally fill the holes in the valence bands, making nonradiative recombination of electrons and holes without photons' emission. Figure 3B shows that the number of excited electrons starts to drop rapidly at 210 fs, indicating the electron-hole nonradiative recombination that releases the atomic driving forces exerting on the Ir-Ir dimers. At this moment, these Ir atoms have been accelerated to high speeds.

These residual speeds drive them towards the nearest neighbor of non-dimerized Ir atoms to form new dimers (Fig. 4). Interestingly, these new dimers are formed even though their lattice temperature ($T \sim 900$ K) is well above the critical temperature $T_S = 280$ K of thermal-induced phase transition[36]. It shows, once again, the phase transition is a coherent kinetic process, not a thermodynamic random process. Although the atomic kinetic energy is important (as demonstrated in the NVE simulation) in a photoinduced phase transition, its role is unlikely as in the thermodynamic phase transition but a kinetic movement to overcome barriers towards breaking the old dimer and forming the new dimers. Accompanying the new Ir-Ir dimerization, the vanished antibonding states of Ir-Ir dimers above the Fermi level re-appear in the PDOS (see Fig. S3). Therefore, we have developed a microscopic explanation for the observed ultrafast recovery of PLD in experiments [7]. We also record the structural phase change evolution dynamics following photoexcitation in a movie (given in SI). In short, we reproduced the experimentally measured change of the diffraction intensity following the photoexcitation, in both magnitude and time scale. All this is only possible after we take into account the carrier cooling effect.

Our excellent agreement with the experiment also supports a fast sub-picosecond energy transfer from electron to phonon. It contrasts with a common perception that the electron to phonon energy transfers is a slow process in several ps, as judged by the observed electron-phonon equilibration time (several ps). [39,40] We believe this long electron-phonon equilibration time is mainly due to the equilibration process of lattice thermal motions following the initial coherent motions with atomic kinetic energy obtained from the carrier cooling. After the lattice vibration reaches equilibrium, the system will finally transfer from the new PLD structure to the HT phase due to its high lattice temperature ($T \sim 900$ K, which is well above the critical temperature $T_S = 280$ K). We indeed notice that a much slower phase transition following photoexcitation has also been reported experimentally in IrTe$_2$.[8] However, this should be distinguished from the sub-picosecond diffraction measurement shown in Fig. 1A. To examine the temperature-induced structural phase transition, we utilize a Born-Oppenheimer MD (BOMD) to simulate the lattice dynamics of the system in the ground state at lattice temperature $T \approx 1000$ K but without photoexcitation (Fig. S4). We indeed find that the Ir-Ir dimers undergo a dissolution at about 6.5 ps, a time scale nearly one order of magnitude slower than the PIPT. In this case, high lattice temperature is the only aspect responsible for this structural phase transition.

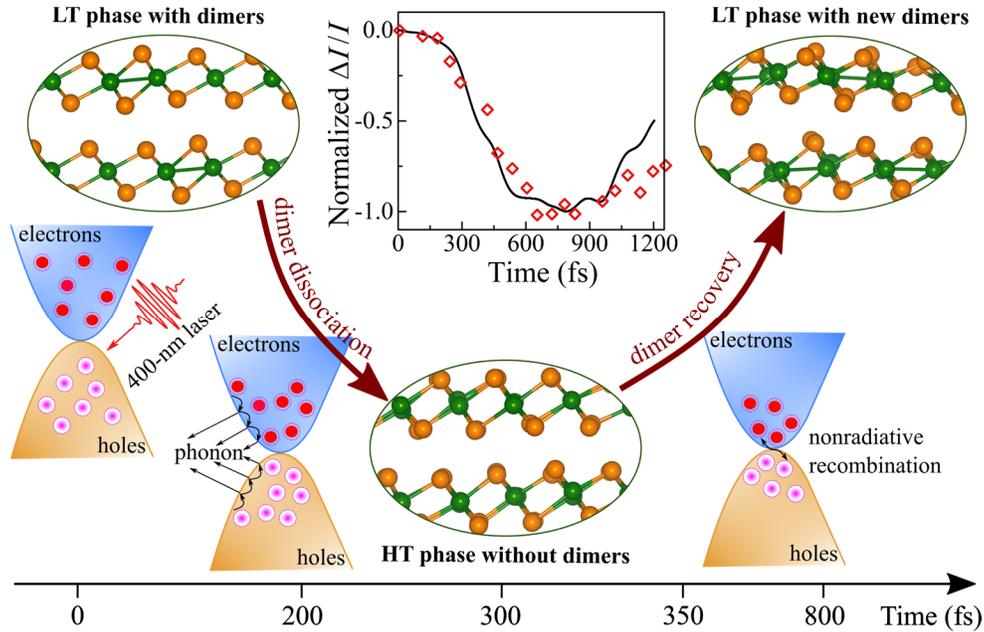

**Fig. 4. Schematic diagram of photoinduced ultrafast dynamics.** The photoinduced valence electrons within 120 fs, producing the electron-hole pairs. The structural dynamics following photoexcitation are driven by electron-phonon couplings.

In summary, we have unraveled the critical role of the hot carrier cooling in the photoexcitation-induced structural phase transition by performing rt-TDDFT simulations including a Boltzmann factor to model the hot carrier cooling process. As summarized in Figure 4, if the pump laser pulse's photon energy is much larger, the photoexcited electrons and holes have a widespread distribution in energy inside the conduction and valence bands. The widespread distribution of photoexcited carriers will not exert a sufficiently large atomic force to dissociate the Ir-Ir dimers. On the other hand, the hot carrier cooling will have two major effects, which help the Ir-Ir dimer dissolution: first, there will be more occupation of the Ir-Ir dimer antibonding state, which enhances the atomic force to dissociate the dimer; second, the atoms of the Ir-Ir dimers will gain considerable coherent kinetic energies, which can help them to overcome any barrier during the dimer breaking process. In a combination of these two factors, the hot carrier cooling yields a phase transition curve in excellent agreement with the experimentally observed time-resolved diffraction data. We further show that the transition is deterministic and coherent. The coherent kinetic energy induces the formation of the new Ir-Ir dimers (at different locations from the initial ones), contributing to the experimentally observed sub-picosecond recovery of the LT phase. On the other hand, for a thermal equilibrium system to induce phase transition due to its high-temperature effect, it can take about 6.5 ps, ten times longer than the fast coherent phase transition caused by hot carrier cooling. We believe our current understanding presents a new insight into photoexcitation-induced ultrafast phase transitions. It is likely also applicable to other ultrafast phase transitions for systems like vanadium dimers in $VO_2$ [9,18] and CDW material 1T-$TaS_2$ [1], 1T-$TiSe_2$ [2,14], and 1T-$LaTe_3$. [13,15]

**Methods**

**Real-Time TDDFT:** We carry out the real-time time-dependent density functional theory (rt-TDDFT) simulations [41] based on the norm-conserving pseudopotentials (NCPP) [42] and Perdew-Burke-Ernzerhof (PBE) functional within density functional theory (DFT) framework with a plane wave nonlocal pseudopotential Hamiltonian, which is implemented in the code PWmat [43]. The wave functions were expanded on a plane-wave basis with an energy cutoff of 45 Ryd. In the rt-TDDFT simulation, we use a 128-atom supercell for $IrTe_2$ and the $\Gamma$ point was used to sample the Brillouin zone. In this rt-TDDFT algorithm, the time-dependent wave functions, $\psi_i(t)$, are expanded by the adiabatic eigenstates, $\phi_l(t)$:

$$\psi_i(t) = \sum_l C_{i,l}(t)\, \phi_l(t) \tag{1}$$

and

$$H(t)\phi_l(t) \equiv \varepsilon_l(t)\phi_l(t) \tag{2}$$

Here, $H(t) \equiv H(t, R(t), \rho(t))$, $R(t)$ represents the nuclear positions, and $\rho(t)$ represents the charge density. By using equation (1), the evolution of the wave functions $\psi_i(t)$ is changed to the evolution of the coefficient $C_{i,l}(t)$. In equation (2), a linear-time-dependent Hamiltonian (LTDH) is applied to represent the time dependence of the Hamiltonian within a time step. Thus, we can obtain a much larger time step (0.1 fs - 0.2 fs) than the conventional real-time TDDFT (sub-attosecond), and the time step is set to 0.1 fs in our simulation.

To mimic the photoexcitation, we apply an external electric field to simulate a laser pulse with a Gaussian shape in our rt-TDDFT,

$$E(t) = E_0 \cos(\omega t) \exp[-(t-t_0)^2/(2\sigma^2)] \qquad (3)$$

where $E_0$ = 0.2 V/Å, $t_0$ = 60 fs, $2\sigma$ = 25 fs is the pulse width, and ω = 3.1 eV is the photon energy. [7]

**Boltzmann Formula in rt-TDDFT:**

To depict hot carrier cooling, we utilize the Boltzmann formula in rt-TDDFT. [30] Boltzmann formula will depend on adiabatic state $\phi_l(t)$ in eq. 1. After Boltzmann correlation ($e^{-t/\tau_{i,j}}$, here, $\tau_{i,j}$ is a decoherence time of 20 fs), a change $\Delta C_{i,l}(t)$ is introduced in the coefficient $C_{i,l}(t)$, thus, the wave function is updated

$$\psi_i(t) = \sum_l (C_{i,l}(t) + \Delta C_{i,l}(t))\phi_l(t) \qquad (4)$$

Compared with total energy ($E_{tot1}$) calculated in eq. 1, the total energy ($E_{tot2}$) is indeed changed in eq. 4. Based on the change of energy, we can obtain the increased kinetic energy which embodies the *el-ph* coupling effect. This energy will be added to the transition degree of freedom in the NVE ensemble. For a more detailed description of the Boltzmann factor formalism, we refer to it in Ref. 30. In the NVT ensemble with carrier cooling, the total kinetic energy of all the atoms is kept the same by a simple rescaling.

**Born-Oppenheimer MD:** A 128-atom supercell for IrTe$_2$ and the $\Gamma$ point to sample the Brillouin zone are used in Born-Oppenheimer MD (BOMD). After geometry optimization at static calculation, we use NVE ensemble: the lattice temperature always oscillates around the set temperature of 1000 K in Fig. S4.

**Acknowledgments:**

The work in China was supported by the Key Research Program of Frontier Sciences，CAS under Grant No. ZDBS-LY-JSC019, the Strategic Priority Research Program of the Chinese Academy of Sciences under Grant No. XDB43020000, and the National Natural Science Foundation of China (NSFC) under Grant Nos. 11925407 and 61927901. L.W.W. was supported by the Director, Office of Science (SC), Basic Energy Science (BES), Materials Science and Engineering Division (MSED), of the US Department of Energy (DOE) under Contract No.DE-AC02-05CH11231 through the Materials Theory program (KC2301).

Supplemental Materials for

# The critical role of hot carrier cooling in optically excited structural transitions

Wen-Hao Liu, Jun-Wei Luo*, Shu-Shen Li, and Lin-Wang Wang*

*Email: jwluo@semi.ac.cn; lwwang@lbl.gov

**This PDF file includes:**

Section 1. Dynamics of photoinduced structural phase transitions.

    Fig. S1. Root-mean-square displacement (RMSD).

    Fig. S2. Comparison of NVE and NVT simulations.

    Fig. S3. Evolution of dimer and non-dimer partial density of state (PDOS).

Section 2. Dynamics of thermally induced structural phase transitions.

    Fig. S4. Thermally-induced structural transition.

**Section 1: Dynamics of photoinduced structural phase transitions**

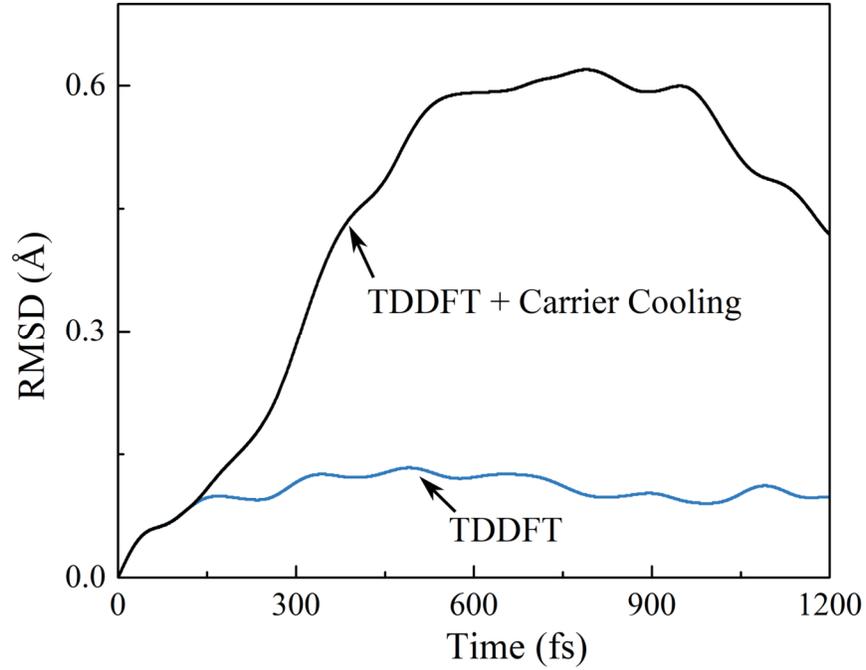

**Fig. S1. The averaged root-mean-square displacement (RMSD) $<u^2(t)>$ of IrTe$_2$ following photoexcitation.** The blue line shows the result of TDDFT without carrier cooling and the black line for the result of TDDFT considering the carrier cooling effect. From averaged RMSD $<u^2(t)>$, we can calculate the diffraction intensity $I(t)$ according to the Debye-Waller formula, $I(t) = \exp[-Q^2<u^2(t)>/3]$, where $Q$ is the magnitude of the reciprocal lattice vector for the reflection probed.

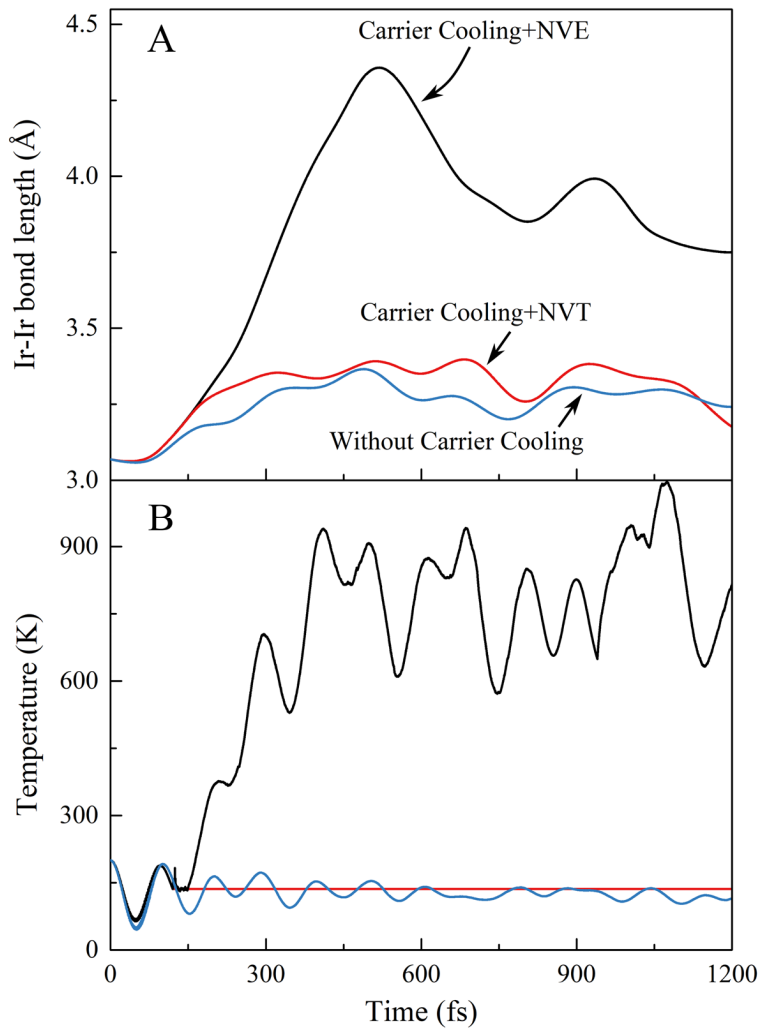

**Fig. S2. Dynamic evolution of Ir-Ir dimers following photoexcitation in different TDDFT simulations.** (A) Evolution of bond length of the Ir-Ir dimers. (B) lattice temperature. The blue line represents the result of the TDDFT simulation without considering hot carrier cooling, and TDDFT simulations considering hot carrier cooling are indicated by the red line for the NVT ensemble and the back line for the NVE ensemble.

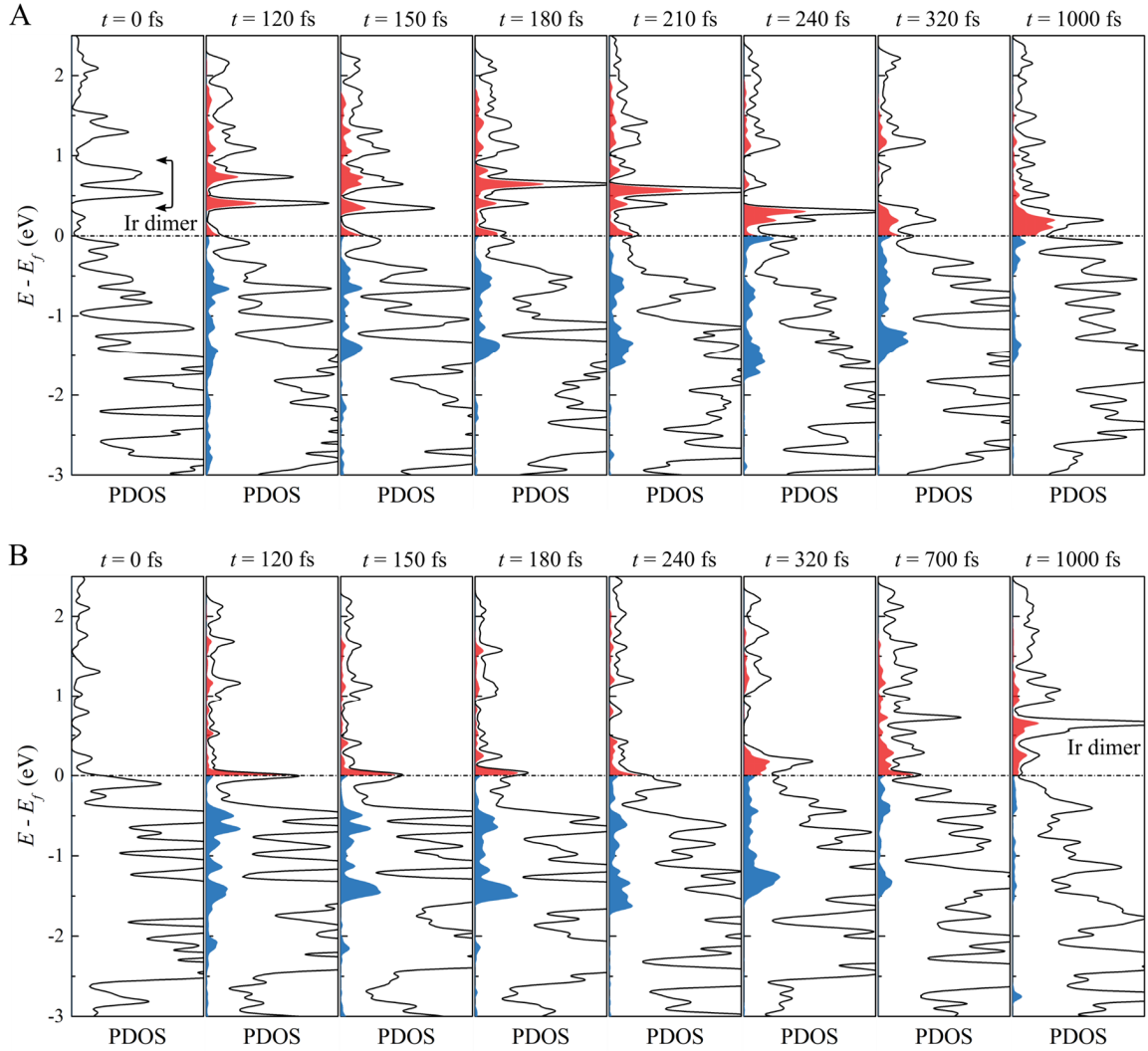

**Fig. S3. Evolution of partial density of state (PDOS) and occupation numbers in IrTe$_2$ following photoexcitation.** (A) Evolution of Ir $d_{xz} + d_{yz}$ PDOS of the Ir-Ir dimers, which undergoes a photoinduced dissolution at 300 fs. (B) Evolution of Ir $d_{xz} + d_{yz}$ PDOS of the non-dimerized Ir atoms, which undergoes a dimerization. Red (blue) shaded areas show the excited electrons (holes). One can see that the Ir-Ir dimerized antibonding state completely disappears at 320 fs and the original non-dimerized Ir atoms start to dimerize after 700 fs.

**Section 2. Dynamics of thermally induced structural phase transitions**

The hot carrier cooling following the photoexcitation heats the lattice subsystem to above the critical temperature for temperature-induced structural phase transition in IrTe$_2$. It is interesting to investigate the thermal effect on photoinduced structural phase transitions. To study the temperature-induced structural phase transitions, we utilize Born-Oppenheimer MD (BOMD) to simulate the dynamics of phase transition at lattice temperature $T \approx 1000$ K as shown in Fig. S7(A). One can see in Fig. S7(B) that the LT-to-HT phase transition occurs at 6.5 ps as characterized by the transferring of the 1/5 periodic charge-modulation structure (LT phase) to an undistorted lattice (HT phase). Note that thermally driven phase transition within the several-picosecond timescale belongs to disordering dynamics. This is nearly 10 times slower than the coherent dissociation of Ir-Ir dimers following photoexcitation.

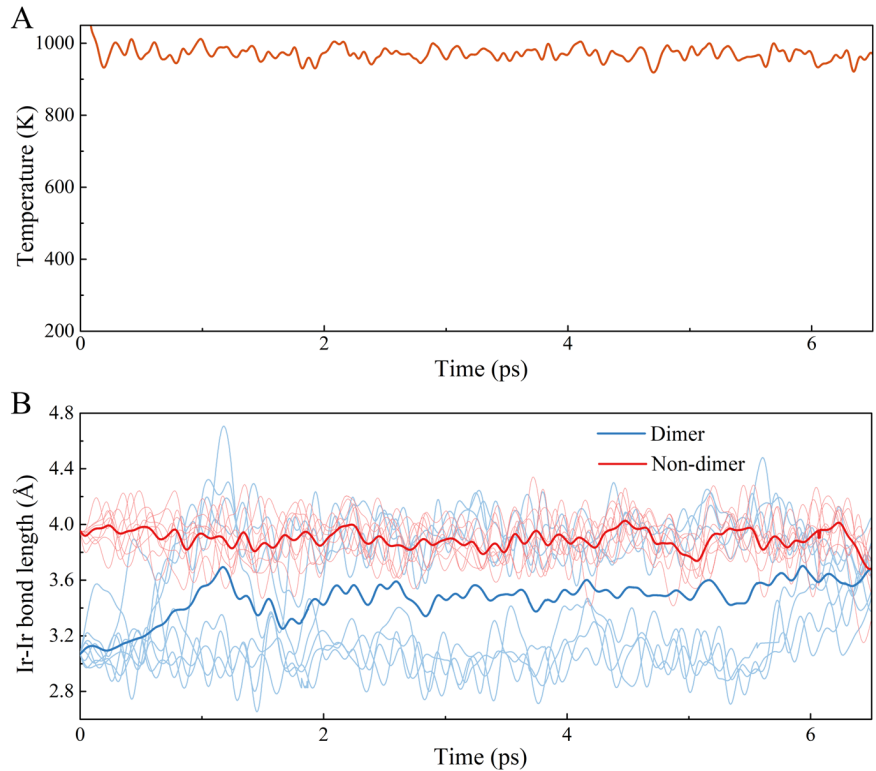

**Fig. S4. Thermally-induced structural transition.** (A) Temperature as a function of time. (B) Evolution of the Ir-Ir dimers (blue lines) and non-dimerized Ir-Ir pairs (red lines).